\documentclass[aps,twocolumn,showpacs,floats,superscriptaddress]{revtex4}
\usepackage{graphicx}
\usepackage{amssymb,amsmath}
\usepackage{color}
\usepackage{dcolumn}
\usepackage{bm}

\def\he4{$^4$He}
\def\Am3{\AA$^{-3}$}
\def\beq{\begin{equation}}
\def\eeq{\end{equation}}
\begin{document}


\author{D. Aleinikava}
\affiliation{Department of Engineering Science and Physics,
CUNY, Staten Island, NY 10314, USA}

\author{A.B. Kuklov}
\affiliation{Department of Engineering Science and Physics,
CUNY, Staten Island, NY 10314, USA}


\title{Stress induced dislocation roughening -- phase transition in 1d at finite temperature} 

\date{\today}

\begin{abstract}
We present an example of a generically forbidden phase transition in 1d at finite temperature -- stress induced and thermally assisted roughening of a superclimbing  dislocation in a Peierls potential. We also argue that such roughening is behind the strong suppression of the superflow through solid \he4 in a narrow temperature range recently observed by Ray and Hallock (Phys.Rev. Lett. {\bf 105}, 145301 (2010)).  
\end{abstract}



\maketitle
Strong interest in the supersolid state of matter in free space \cite{SFS} has been revived by the recent discovery of the torsional oscillator (TO) anomaly in solid \he4 \cite{KC}. While finding no supersolidity in the ideal \he4 crystal, {\it ab initio} quantum Monte Carlo simulations did find that some grain boundaries \cite{GB}, dislocations \cite{screw,sclimb} or crystal boundaries \cite{Ceperley} support low-d superfluidity spatially modulated by the surrounding lattice. In principle, a percolating network of superfluid dislocations \cite{Shevchenko} could explain the TO anomaly if the dislocation density is 3-4 orders of magnitude higher than it is expected to be in a slowly grown and well annealed crystal.
Consistent with such expectation is also a very small rate of the critical superflow through solid \he4 (occuring presumably along dislocations with superfluid cores) observed in the UMASS-Sandwich experiments \cite{Ray,Hallock_2010}. Thus the nature of the TO anomaly in solid \he4 remains unclear.

In the present work we focus on the very unexpected feature of the UMASS-Sandwich experiment \cite{Hallock_2010} -- the strong suppression of the supercritical flow rate $V_{cr}$ (by about 3-4 times!) and then its  recovery in a  narrow range of temperatures. Such a feature occurs well below (about 10 times) the flow onset temperature $T_O \approx 0.5-0.6$K \cite{Hallock_2010}. 
Here we are proposing an explanation within the model of superclimbing dislocation \cite{sclimb}, biased by an externally imposed chemical potential $\mu$ which generates  a stress on the dislocation core and, thus creates spontaneous jog-antijog pairs. 

Jog-antijog pairs as quantum objects can be created spontaneously by a {\it macroscopically} small stress $\sigma \geq \sigma_c\propto 1/L$ applied to a superclimbing dislocation of length $L$ -- analogous to the creation of kink-antikink pairs along a stressed gliding dislocation \cite{Hirth,Petukhov}. We have found that such an instability leads to a first-order phase transition even at finite temperature $T$ between two phases of the dislocation -- {\it smooth} and {\it rough}. This transition is in an apparent violation of Landau's argument "no phase transitions in 1d at finite $T$" \cite{Landau}. However, we argue that the locality of order parameter(s), which is essential for the validity of Landau's argument, is not present here. Consequently, in a sharp contrast with conventional 1d systems, where any macroscopic characteristic length {\it decreases} with increasing $T$ \cite{Landau}, a typical scale $L_h$ for the onset of the hysteretic behavior of the superclimbing dislocation {\it increases} with $T$.

The effective description of such a transition invokes a  single coarse grained macroscopic degree of freedom -- dislocation deformation characterized by an effective mass and a potential energy with two minima. These quantities are scaled as some positive powers of $L$ even at $T \neq 0$, so that  the amplitude of the transition between the minima decays exponentially  as $L\to \infty $ -- very much like $d>1$ systems undergoing first-order transition. However, due to the strongly interacting and effectively long-range nature  of the {\it rough} phase,  specifics of such size dependencies cannot be derived analytically, and we have evaluated them numerically.   
\\

\noindent{\it The model and its Monte-Carlo simulations}. Superclimbing dislocation is modeled as a quantum string oriented along the $x$-axis and strongly pinned at its both ends $x=0,L$ \cite{Granato}. The string displacement $y(x,t)$ along  the $y$-axis depends on the time $t$ and is measured in units of the  inter-atomic spacing ($\approx$ Burger's vector $b$) with respect to its equilibrium $y=0$ (no tilting is considered). The Peierls potential induced by the crystal is taken as $U_P=-u_P \cos\left(2\pi y(x,t)\right)$. 
The partition function $Z$ has the form \cite{sclimb,JLTP}   
\begin{eqnarray} 
Z&=&\int Dy(x,t)\, D\rho(x,t) D\phi(x,t) \exp(-S), 
\label{Z} \\
S&=& \int_0^\beta dt\sum_{x}[ i(\rho +n_0)\nabla_t \phi  + \frac{\rho_0}{2} (\nabla_x\phi)^2  
\nonumber \\
&+&\frac{1}{2\rho_0} (\rho - y)^2 + {m\over{2}} \left((\nabla_t y)^2 + V_d^2(\nabla_x y)^2 \right)  
 \nonumber \\
&-&  u_P \cos\left(2\pi y(x,t)\right) -F y(x,t)] \label{H},
\end{eqnarray}
 where all the variables are periodic in the imaginary time $t\geq 0$ with the period $\beta=1/T$ (units $\hbar=1$, $K_B=1$); the core density $ \rho$ and the superfluid phase $\phi$ are canonically conjugate variables, with
$\rho'=\rho -y$ being the local superfluid density;
the derivatives $\nabla_{t,x} y$, $\nabla_{t,x} \phi$ are understood as finite differences in the discretized space-time lattice (with 200 time slices and
$x=0,1,2,...,L$ in units of $b$), with 
$\nabla_{t,x} \phi$ defined modulo 2$\pi$ (in order to take into account phase-slips); $n_0, \rho_0$ stand for the average filling factor ( we choose $n_0=1$) and the bare superfluid stiffness, respectively, with the bare speed of first sound taken as unity.

The first two terms in Eq.(\ref{H}) describe the superfluid response of the core \cite{sclimb}, and  the third term accounts for the superclimb effect \cite{sclimb} -- building  the dislocation edge so that the core climb $y\to y \pm 1, \pm 2, ...$ becomes possible by delivering matter $\rho \to \rho \pm 1,\pm 2, ...$, respectively, along the core \cite{sclimb}. The dislocation is assumed to be attached to large superfluid reservoirs at both ends, with spatially periodic boundary conditions for the supercurrent.

The terms $\propto m $ in Eq.(\ref{H}) account for the elastic response of the string, with $m$ and $V_d$ standing for the  effective mass of the dislocation core (per $b$) and the bare speed of sound, respectively. Since the main source of kinetic energy are supercurrents, we have left out the
term $\sim (\nabla_t y)^2$ in Eq.(\ref{H}). 
The parameter $m$ in Eq.(\ref{H}) is not actually a constant.  It contains a contribution from the Coulomb-type interaction potential $ \propto 1/|x|$ between jogs (or kinks, cf. \cite{EPL}) separated by a distance $x$ \cite{Hirth}. Accordingly, $m$ has a logarithmic divergent factor 
  with respect to a wave vector $q$ along the core $m(q) =  m_0\cdot\left[1+ U_C \ln\left(1+ \frac{1}{(bq)^2}\right)\right]$, where $m_0$ is of the order of the atomic \he4 mass and $U_C \sim 1$ is a parameter characterizing the strength of the interaction \cite{EPL,JLTP}. 
  In solid \he4, the zero-point fluctuation parameter $K=\pi \hbar/(4m_0 bV_d) \sim 1$  \cite{EPL}. We present our numerical results for $U_C=1, V^2_d=5, K=1$. 
It is important to note that the main results are not qualitatively  sensitive to the long-range interaction. 

The linear force density $F\approx b\sigma$ (ignoring spatial indices) in Eq.(\ref{H}) is determined by 
 the external stress $\sigma$ induced by the chemical potential
difference $\delta \mu$ applied in the setup \cite{Hallock_2010}: $\delta \mu \approx \delta p b^3$, with $\delta p$ being the resulting overpressure. Thus $\sigma = \delta \mu/b^3$, i.e. in units of $b$ $\sigma=\delta \mu=\delta p=F$.

Monte Carlo simulations  have been conducted with the Worm Algorithm (WA) \cite{WA} for the superfluid part of the action, with the Peierls term treated within the Villain approximation similarly to Refs. \cite{EPL,JLTP}.
The  renormalized superfluid stiffness $\rho_s(T,F)$ and compressibility $\kappa(T,F)$ have been calculated in terms of the windings of the dual variables \cite{WA,EPL,JLTP}. 
 No significant effect of the bias $F$ was found on $\rho_s(T,F)$, and thus we consider $\rho_s(T,F)=\rho_s(T,0)\equiv \rho_s(T)$. In contrast, $ \kappa(T,F)$ does experience quite dramatic renormalization, which is the focus of the present work.

The bare stiffness $\rho_0(T)$ vanishes above some temperature $T_0$ comparable to the bulk $\lambda$-temperature ($\sim 1-2$K). We have used $T_0=0.2$ (in the dimensionless units as a fraction of the Debye temperature $T_D$ for the first sound), and considered low temperatures --  such that $\rho_s(T)$ stayed unchanged within 1-10\% of its $T=0$ value.  In other words, the thermal length $L_T \approx \rho_s(0)/T $, above which $\rho_s(T)$ becomes suppressed, is the largest scale in the problem.

We have also calculated full $\chi_1 = \delta N /\delta F $ and differential $\chi_2=dN/dF$ 
isochoric compressibilities of the dislocation \cite{JLTP}, where
$\delta N$ is the full amount of matter accumulated in the extra plane forming the superclimbing (edge) dislocation in response to the variation of $F$.
These compressibilities obey the relationship $\chi_2= \chi_1 + Fd\chi_1/dF$, and in the linear regime $\chi_1 \approx \chi_2$. It is found that $\kappa \propto \chi_2/L$ in the strongly fluctuating regime. However, despite such similarity, $\kappa$ and $\chi_2/L$ are not one and the same quantity: while $\chi_2$ describes the shifting of the dislocation position, $\kappa$ accounts for the time dependent response of the superfluid phase $\phi$. In other words, $\kappa$ and $\rho_s$ enter as renormalized coefficients of the effective superfluid action $S_\phi= \int dx \int d\tau [ \rho_s (\partial_x \phi)^2 /2 + \kappa \, (\partial_t \phi)^2/2]$ \cite{Haldane}, where the renormalized speed of first sound
\begin{equation}
V_s(T,F)=\sqrt{\rho_s/\kappa}
\label{Vsf} .
\end{equation}

In the absence of the Peierls potential or at high $T$ (where still $\rho_s(T)\approx \rho_s(0)$),  $\chi_{1,2}$ are practically equal to the free string response $\chi_0 \approx  L^3/[12 m_0 (1 +2U_C \ln (L/b)) V^2_d]$ (considered in Ref. \cite{Granato} for $U_C=0$). We introduce the normalized quantities $R_{1,2}\equiv \chi_{1,2}/\chi_0$:
\begin{eqnarray}
R_2= R_1+ FdR_1/dF, \label{R2_kappa} 
\end{eqnarray}
so that $R_2=R_1=1$ for $u_P=0$. 

Our main findings presented below are the following: i) a narrow dip in $V_s(T,F)$ vs $T$ at some {\it macroscopically} small $F=F_c$; ii) periodicity of  $V_s(T,F)$ with respect to the bias $F$; iii) exponential scaling  of the dip depth with $L$; iv) hysteresis developing beyond a certain length $L_h$ {\it growing} with $T$.\\
\begin{figure}
\centerline{\includegraphics[angle =0,width=0.9\columnwidth]{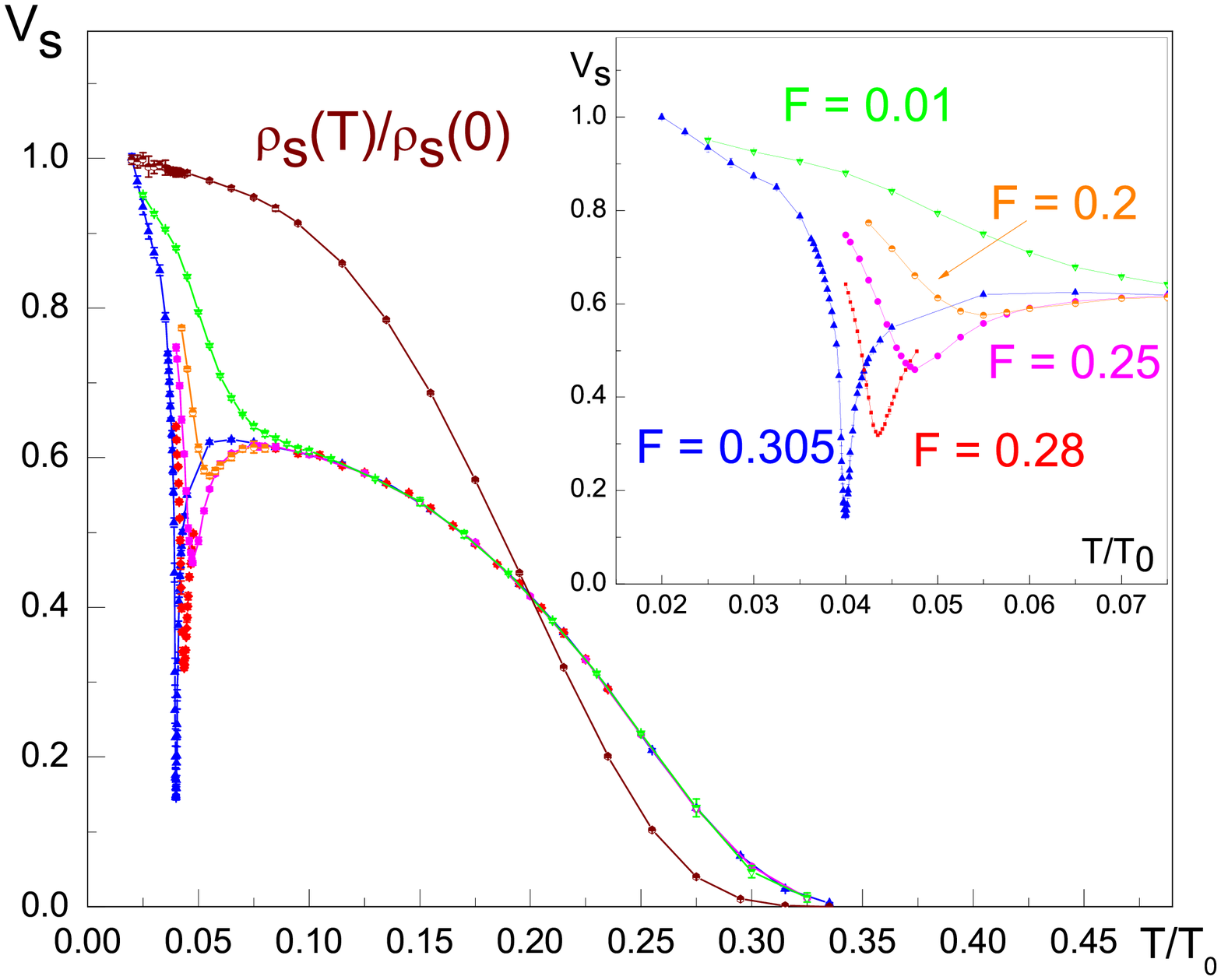}}
\vspace{-0.5cm}
\caption{(Color online) Renormalized superfluid stiffness $\rho_s(T)$ and the velocity $V_s(T,F)$ of first sound  normalized by their respective low-$T$ values for different 
$F $ (shown on the inset), $L=30$, $u_P=3.0$. Inset: the region of the dip (cf. Fig.4 of Ref.\cite{Hallock_2010}) showing its shifting with $F$.}\label{VsT}
\end{figure}
\begin{figure}
\centerline{\includegraphics[angle =0,width=0.9\columnwidth]{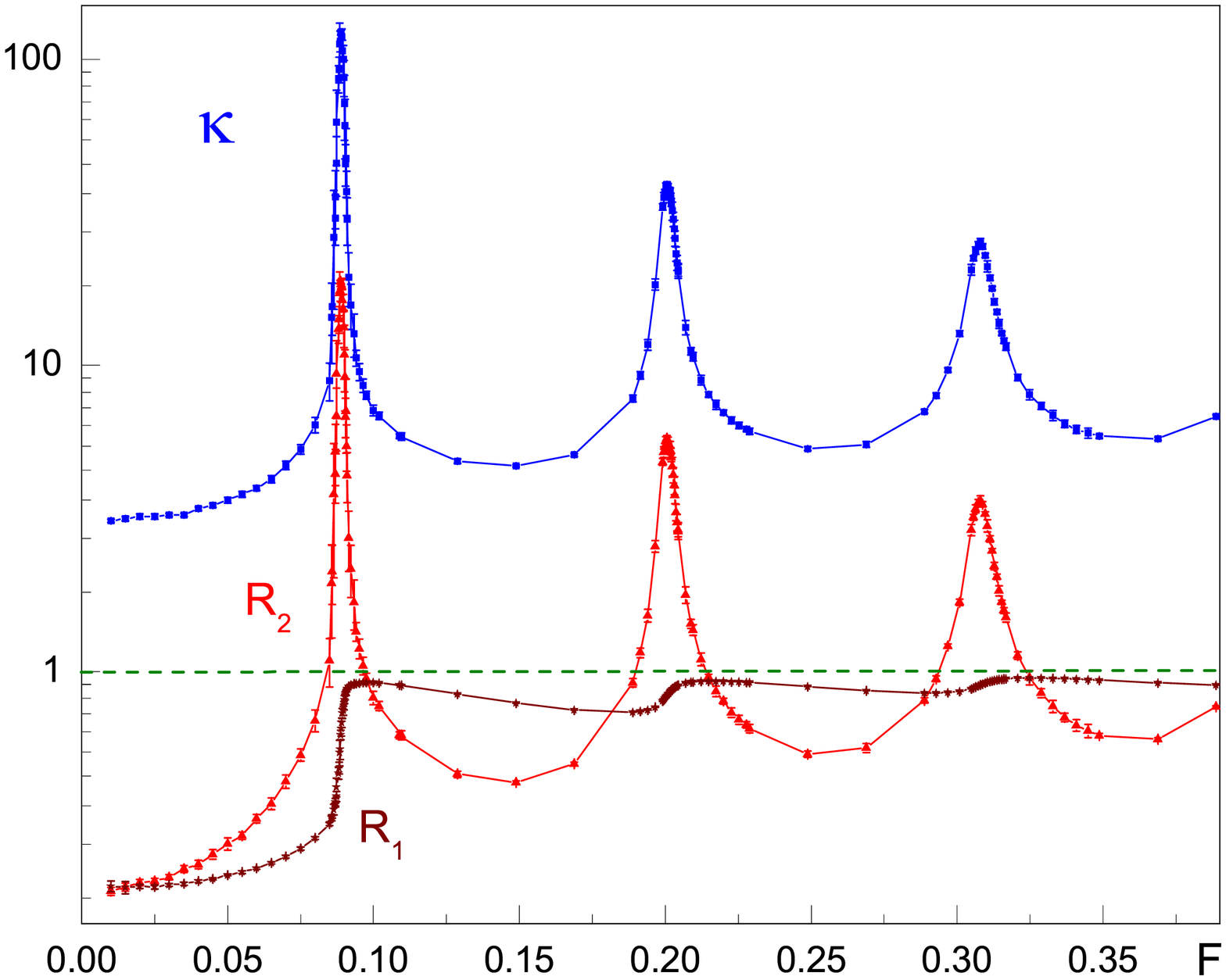}}
\vspace{-0.5cm}
\caption{(Color online) Typical behaviors of $\kappa, \, R_{1,2}$ vs $F$: $L=56, \, T/T_0=0.05, u_P=3.0$. 
Dashed line -- the prediction of the free string model ($u_P=0$) \cite{Granato}.}\label{multi}
\end{figure}
\begin{figure}
\centerline{\includegraphics[angle =0,width=0.9\columnwidth]{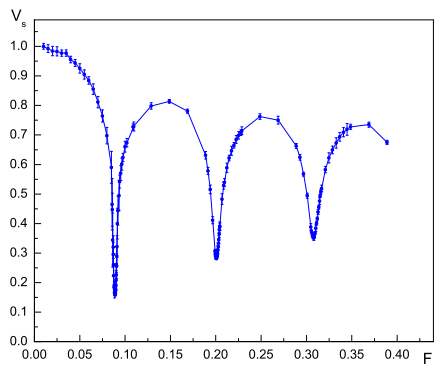}}
\vspace{-0.5cm}
\caption{(Color online)  $V_s(F) $ normalized as in Fig.\ref{VsT}. The parameters are the same as in Fig.\ref{multi}. The narrow dips occur at the thresholds $F=F_c(L,n),\, n=1,2,3$ for $n$ jog-antijog pairs creation (see the text -- {\it Periodicity vs external bias}).}\label{VsF}
\end{figure}
\begin{figure}
\centerline{\includegraphics[angle =0,width=0.9\columnwidth]{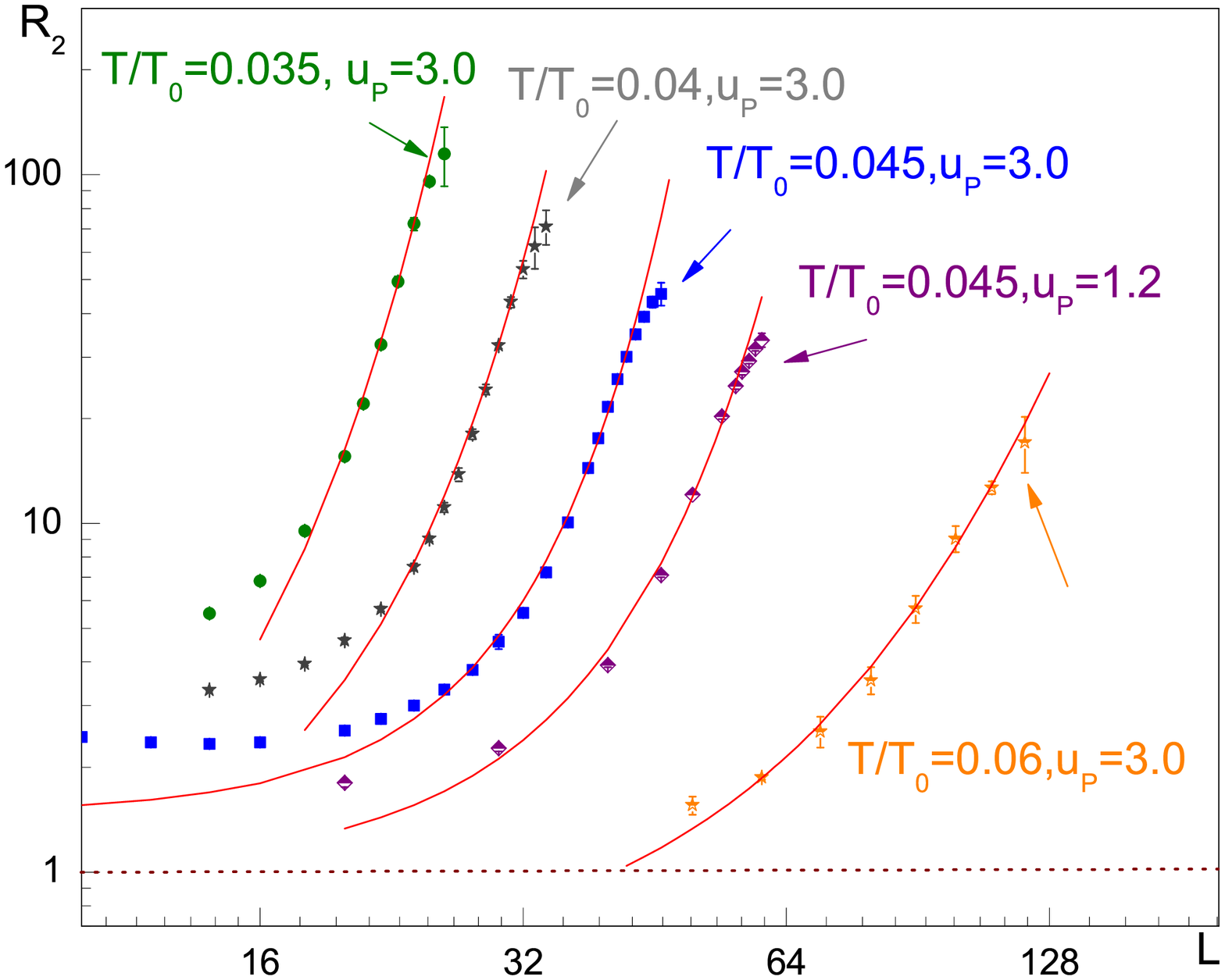}}
\vspace{-0.5cm}
\caption{(Color online) Height of the first resonance peak in $R_2(F=F_c)\propto \kappa $ vs  $L$ for various parameters (symbols) and its fit (lines) by $R_2=\exp(A+L/L_R)+B$ with three adjustable parameters $A,L_R,B$.  Deviations from the fit at large $L$ (shown by arrows) mark the beginning of the hysteretic behavior. Dashed line --  the free string model value ($R_1=R_2=1$) \cite{Granato}. }\label{R2}
\end{figure}
\begin{figure}
\centerline{\includegraphics[angle =0,width=0.9\columnwidth]{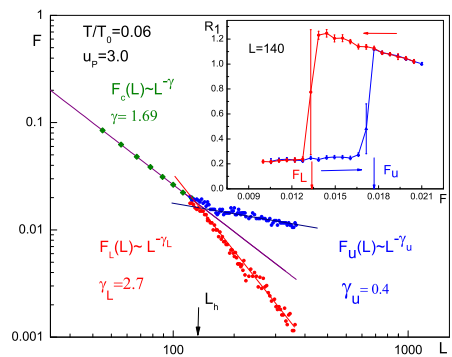}}
\vspace{-0.5cm}
\caption{(Color online) 
Plots of $F_c,\, F_u,\, F_L$ vs $L$ (symbols) and their fits (lines). The vertical arrow separates the resonant-type  peaks in Fig.\ref{multi} ( $L<L_h$), and the hysteresis ($L>L_h$). Inset: a typical  scan in $F$ shown by the horizontal arrows, with the vertical arrows indicating the upper $F_u$ and the lower $F_L$ "coercivity" stresses.}\label{Hyst}
\end{figure}

\noindent{\it Dip in the flow rate}.  
The speed (\ref{Vsf}) sets the value of the critical flow rate $V_{cr}\approx V_s(T,F)$  as a  threshold for generating phase slips in 1d -- similar to the scenario of Ref.\cite{Kagan}.   
Since $\kappa$ (and $R_2$) exhibits a resonant-type peak, $V_s$ acquires the dip shown in Fig.\ref{VsT}. Its depth depends on $L$  and how closely $F$ is tuned to the threshold value $F_{c}\propto L^{-\gamma},\, \gamma \approx 1-1.7$, for the jog-antijog pair creation at given $T$ (see the inset,Fig.~\ref{VsT}). We believe such a dip has actually been detected by Ray \& Hallock (cf. Fig.4 in Ref.\cite{Hallock_2010}).  Away from the dip, $V_s$ is practically insensitive to  $F$, and the responses $\kappa,\, R_2$ become essentially linear in $F$ \cite{JLTP}. \\

\noindent {\it Periodicity vs external bias}. One of the most striking features we have found is the quasi-periodicity in $F$ of $\kappa$, $R_2$ as shown in Fig.~\ref{multi}, 
and in $V_s$ as shown in Fig.~\ref{VsF}. We attribute this to the reaching of the thresholds for creating multiple jog-antijog pairs. Thus, the peak (dip) positions are given as $ F=F_c(L,n) \approx n F_c(L,1)\sim n L^{-\gamma} ,\,\, n=1,2,3,...$, and we predict that the dip in the flow rate observed in Ref.\cite{Hallock_2010} (see Fig.4 there) should recur as a function of the applied bias, provided $T$ is kept fixed. \\
 
\noindent{\it Size dependencies}.
Transformation between the {\it smooth} (where $R_1<<1$) and the {\it rough} ($R_1 \approx 1$) states occurs within an exponentially narrow region $\delta F$ around $F_c$.
It is given by the tunneling rate $ \sim \exp(-L/L_R) << 1$ through the macroscopic jog-antijog barrier (cf. the mechanism for the kink-antikink tunneling, Ref.\cite{Petukhov}), where   $L_R$ stands for the tunneling length.
 Thus, as follows from Eq.(\ref{R2_kappa}), the peak value of $R_2(T,F_c)$ diverges with $L$ as $ \approx 1/\delta F \sim \exp(L/L_R)$. 
Fitting $R_2(T,F_c)$ by an exponential function, Fig.\ref{R2}, gives $L^{-1}_R \approx L^{-1}_0 (1-T/T_R)^{2.3}$. Here $L_0$ is the $T=0$ tunneling length ($L_0 \approx 1.7$ for $u_P=3.0$) and $T_R $ sets the scale for thermal roughening, i.e.
the temperature above which the density of  jog pairs is large even in the limit $F\to 0$. $T_R$  is determined by the double energy $2\Delta \propto \sqrt{u_P}$  of a jog modeled as a Sine-Gordon soliton. We have found $ T_R \propto u_P^{s}, \, s=0.5 \pm 0.1 $ which is consistent with such an interpretation.

The critical stress $F_c(L,1)$ has been found to deviate from the $1/L$ dependence \cite{Petukhov} at finite $T$. Specifically,  $F_{c} \propto 1/L^{\gamma(T)} $ with $\gamma(T) >1$. As $T$ grows, $\gamma(T) \to 1.7$, and $\gamma(T) \to 1$ in the limit $T\to 0$. The temperature scale for this variation is set by the Peierls potential amplitude $u_P$ as well. It is natural to attribute the deviation from $\gamma=1$ to a suppression of the energy gap  $ \Delta \propto L^{1-\gamma} \to 0$ at finite $T$.\\

\noindent {\it Hysteresis.} 
The resonant-peak type behavior in $R_2$ (and in $\kappa$) turns out to be a precursor for the jump in $R_1$. As seen from Eq.(\ref{R2_kappa}), $R_2 \approx F_cdR_1/dF \sim F_c/\delta F$ as $\delta F \to 0$. The  hysteresis emerges when $\delta F$ becomes significantly less than the coexistence region  for the {\it smooth} and {\it rough} states. This conditions sets a typical length $L_h$ above which ($L>L_h$) hysteresis develops.
We have found that 
$L_h$ {\it grows} with  $T$ as $ L_h \approx L_0 (T/T_s)^{\gamma_h} \gg L_0,\, \gamma_h>0$ ($\gamma_h \approx 2-3$ for $u_P = 1-3$). We attribute the energy scale $T_s$  to the tunneling splitting energy through the microscopic jog-antijog barrier so that $ T_s \ll T_R$ \cite{note}. This feature is clearly due to the collective multi-jog nature of the {\it rough} state, and it deviates strongly from the single pair tunneling scenario \cite{Petukhov} (see Eq.(39)) where the tunneling rate saturates at some length {\it decreasing} as $ \propto T^{-3/2}$. 

Fig.\ref{Hyst} demonstrates  the $L$-dependencies of the "coercivity fields".
Along the lower branch of the hysteresis loop (inset in Fig.\ref{Hyst}) the dislocation is in the {\it smooth} state. Upon increasing $F$, it "jumps" into the {\it rough} state at $F\approx F_u$. While "moving" back along the upper branch representing the {\it rough} state, the dislocation returns  into the {\it smooth} state at $F\approx F_L < F_u$. 
As seen from the main panel of Fig.\ref{Hyst}, both fields scale as $F_{u,L} \propto L^{-\gamma_{u,L}}$, with $\gamma_u \approx 0.4$ and $\gamma_L\approx 2.7$, respectively. The middle straight line corresponds to the extrapolation of the peak position data $F_c \propto L^{-\gamma}, \, n=1,\, \gamma \approx 1.7$ (for $L<L_h$).
Such strong sensitivity to the size $L$ as well as $L_h$ growing with $T$ clearly indicate that 
the stress-induced roughening is a phase transition at finite $T$ in 1d.

 The resonant-type behavior and the hysteresis have also been found in simulations of gliding dislocation \cite{JLTP}.
While $L_T$ determines the upper spatial scale for the superclimb, no such restriction exists for the glide, so that the formal limit $L\to \infty$ can be considered. 
\\

\noindent{\it Discussion and conclusions}. The narrow dip in the superflow rate observed in Ref.\cite{Hallock_2010} may have its origin in the stress induced roughening effect of superclimbing dislocations. Specifically, biasing  a superfluid dislocation network 
by {\it macroscopically} small overpressure 
can induce strong  suppression of the first sound along the superfluid cores. Such suppression is characterized by the periodicity of the dip in the flow rate, Fig.\ref{VsF},  which can be used as  the {\it experimentum crucis} for the proposed scenario.
We estimate the critical overpressure $\delta p $ in terms of  $L$ and a typical jog-antijog energy $2\Delta \approx 0.1$K (cf. \cite{EPL}) as $\delta p/p \approx  2\Delta b/(LT_D) \sim 10^{-2} b/L $. Measurements of $\delta p$ where the dip recurs can provide crucial information on the nature of the dislocation network  --  its typical free segment length $L$. 

As $T$ is lowered, the opposite condition $L>L_h$ is fulfilled so that the dislocation behavior becomes hysteretic between its smooth and rough states. We believe it is  also important to study the hysteresis in the flow rate vs  chemical potential (the upper $F_u$ and the lower $F_L$ fields) at different temperatures.
\\


We are grateful to R.B. Hallock, L.P. Pitaevskii, N.V. Prokof'ev, D. Schmeltzer and B.V. Svistunov for useful discussions and comments. 
This work was supported by  the National Science Foundation, grant No.PHY1005527,  PSC CUNY, grant No. 63071-0041, by the CUNY HPCC under NSF Grants CNS-0855217 and CNS - 0958379.


\end{document}